\documentclass[pra,twocolumn,showpacs,superscriptaddress,floatfix, nofootinbib]{revtex4-1}
\usepackage[caption=false]{subfig}
\usepackage{amsmath}
\usepackage{graphicx}
\usepackage{epsfig}
\usepackage{amssymb}
\usepackage{mathrsfs}
\usepackage{esint}
\usepackage{gensymb}
\usepackage[bookmarks=true,
   colorlinks=true,
   linkcolor=blue,
   urlcolor=blue,
   citecolor=blue,
   bookmarks=true,
   hyperindex=true
]{hyperref}
\usepackage{natbib}
\usepackage{relsize}
\usepackage{float}
\usepackage{dsfont}
\usepackage{multirow}
\usepackage{bm}
\newcommand{\be}{\begin{equation}}
\newcommand{\ee}{\end{equation}}

\newcommand{\beq}{\begin{eqnarray}}
\newcommand{\eeq}{\end{eqnarray}}

\def\H1{\widehat{H}_1}

\usepackage[normalem]{ulem}

\begin{document}

\title{Quantum and quantum-inspired optimization \\ for solving the minimum bin packing problem}

\author{A.A~Bozhedarov}
\affiliation{Russian Quantum Center, Skolkovo, Moscow 143025, Russia}

\author{A.S.~Boev}
\affiliation{Russian Quantum Center, Skolkovo, Moscow 143025, Russia}

\author{S.R.~Usmanov}
\affiliation{Russian Quantum Center, Skolkovo, Moscow 143025, Russia}

\author{G.V.~Salahov}
\affiliation{Russian Quantum Center, Skolkovo, Moscow 143025, Russia}

\author{E.O.~Kiktenko}
\affiliation{Russian Quantum Center, Skolkovo, Moscow 143025, Russia}

\author{A.K.~Fedorov}
\affiliation{Russian Quantum Center, Skolkovo, Moscow 143025, Russia}

\begin{abstract}
Quantum computing devices are believed to be powerful in solving hard computational tasks, in particular, combinatorial optimization problems.
In the present work, we consider a particular type of the minimum bin packing problem, which can be used for solving the problem of filling spent nuclear fuel in deep-repository canisters that is relevant for atomic energy industry. 
We first redefine the aforementioned problem it in terms of quadratic unconstrained binary optimization.
Such a representation is natively compatible with existing quantum annealing devices as well as quantum-inspired algorithms.
We then present the results of the numerical comparison of quantum and quantum-inspired methods. 
Results of our study indicate on the possibility to solve industry-relevant problems of atomic energy industry using quantum and quantum-inspired optimization.
\end{abstract}

\maketitle

\section{Introduction}

Optimization is a primary tool with numerous applications across various industries~\cite{Paschos2014}.
Specific attention is traditionally paid to combinatorial optimization problems, which are especially difficult in the view of the so-called curse of dimensionality --- a dramatic increase of the complexity with increasing problem size.
One of the notable classes of combinatorial optimization problems is quadratic unconstrained binary optimization (QUBO)~\cite{Lucas2014,Fedorov2022}, which appears in various applications.
Quantum computing devices, both universal and specialized, are considered to be useful in solving such computational problems~\cite{Farhi2000,Das2008,Lidar2018,Fedorov2022,Farhi2014}.
An idea behind, generally speaking, is to encode a cost function in a quantum Hamiltonian~\cite{Lucas2014}, so that its low-energy state corresponds to the minimum of the cost function.
Several architectures of quantum computing devices, which are of interests for solving optimization problems, have been developed~\cite{Fedorov2022}.
Specifically, quantum annealing devices based on superconducting qubits, which are able to solve problems of a non-trivial size~\cite{Amin2021}, have been used to tackle various industry-relevant tasks, 
including quantum chemistry calculations~\cite{Leib2019,Chermoshentsev2021}, (lattice) protein folding~\cite{Aspuru-Guzik2012-2,Fingerhuth2018}, genome assembly~\cite{Fedorov2021,Sarkar2021},
solving polynomial~\cite{Sota2019} and linear systems of equations~\cite{Sota2019},
financial optimization~\cite{Orus2019,Orus2020,Grant2021,Alexeev2022,Orus2019-2,Rosenberg2016-2,Rosenberg2016,Rounds2017,Vesely2022},
traffic optimization~\cite{Neukart2017,Inoue2021,Hussain2020},
scheduling~\cite{Venturelli2016,Ikeda2019,Sadhu2020,Botter2020,Domino2021,Domino2021-2}, railway conflict management~\cite{Domino2021,Domino2021-2}, and many others (for a review, see Ref.~\cite{Fedorov2022}).
An alternative approach is to use programmable quantum simulators based on atomic arrays~\cite{Browaeys2020-2}, 
where the most recent advances include a demonstration of a superlinear quantum speedup in finding exact solutions for the hardest maximum independent set graphs~\cite{Lukin2022}.
One may also note that gate-based running variational optimization algorithms, mainly quantum approximate optimization algorithm~\cite{Farhi2014}, also offer interesting possibilities for combinatorial optimization~\cite{Babbush2021,Bharti2021}.
Although such devices in principle are able to demonstrate quantum computational advantage in near future, still various limitations make it challenging to use them for solving problems of industry relevant sizes. 

The problem of a clear comparison between quantum and classical algorithms, which can be used to highlight the quantum origin of the speed up, is also nontrivial~\cite{Troyer2014}.
As a result of such a comparison, a new class of algorithms and techniques, know as {\it quantum-inspired}, has been developed~\cite{Tiunov2019,Lloyd2019}.
As soon as these algorithms are compatible with currently existing (classical) hardware, analyzing their limiting capabilities and advantages over classical approaches are required towards their use in practice. 
Recently, solving the wavelength assignment problem in telecommunication using quantum-inspired algorithm SimCIM~\cite{Tiunov2019} has been demonstrated~\cite{Boev2022}.
For a wide range of benchmark of quantum-inspired heuristic solvers for quadratic unconstrained binary optimization, 
namely D-Wave Hybrid Solver Service, Toshiba Simulated Bifurcation Machine, Fujitsu Digital Annealer, and simulated annealing, see Ref.~\cite{Oshiyama2022}.

A specific class of a combinatorial optimization problem that appear across many industry application is the minimum bin packing problem,
where items of different sizes must be allocated into a finite number of bins (containers), each of a fixed given capacity, in a way that minimizes the number of bins~\cite{Paschos2014};
this problem is known to be NP-hard.
A particular application of this problem is optimization of spent nuclear fuel (SNF) filling in canisters for the deep repository.
According to existing standards, the deposing should be realized by using special (deep-repository) canisters, so that the maximum heat output per canister does not exceed the limiting value.
The tasks of the optimization of the SNF using canister filling (CF) is then clearly linked to the aforementioned minimum bin packing problem~\cite{Zerovnik2009}.
The use of combinatorial methods to optimize the filling of SNF in metal canisters for the final deep repository, 
according to the maximal allowed thermal power per canister and the limit in the number of spent-fuel assemblies per canister, has been demonstrated~\cite{Zerovnik2009}.
In this context, quantum and quantum-inspired tools are now considered as a way to solve this problem for larger sizes; 
in particular, optimization of fuel arrangements in nuclear power plants using quantum tools has been considered~\cite{Whyte2021}.

In this work, we present a method for solving the SNF management problem using quantum and quantum-inspired annealing.
We first formulate the problem in the QUBO form, which allows solving this problem using various annealing tools, including quantum annealing. 
We then benchmark its solution using quantum annealing device from D-Wave\footnote{The results of the present paper are based on the data that have been collected during the availability of the device.}, 
quantum-inspired algorithm SimCIM~\cite{Tiunov2019}, 
and quantum-inspired Simulated Bifurcation Machine (SBM)\footnote{The results of the present paper are based on the data that have been collected during the availability of the algorithm.}~\cite{Goto2019}.
Our results indicate the possibility to solve such an industry-relevant problem using quantum and quantum-inspired annealing.

Our work is organized as follows. 
In Sec.~\ref{sec:CF}, we formulate the CF optimization problem in the QUBO form, which makes it suitable for solving this using quantum and quantum-inspired annealing.
Sec.~\ref{sec:bench}, describes the numerical analysis setup;
there we also benchmark a solution of the CF problem using available the quantum annealer and quantum-inspired annealing algorithms.
We summarize our results and conclude in Sec.~\ref{sec:conclusion}.

\section{Canister filling optimization problem}\label{sec:CF}

The SNF is a subject of the deposition for further safe keeping.
Existing industrial standards require that the deposing should be realized using special (deep-repository) canisters, so that the total heat output per canister does not exceed the limiting value $P_{\max}$.
At the same time, there is a minimum number of spent fuel elements that can be stored in one canister $N_{\min}$.
This is a subject of the SNF management problem, which is important for the optimal use of existing canisters without violating standards.
The SNF management problem can be formulated as a combinatorial optimization problem as follows.

Let $n$ be the total number of spent fuel elements, $m$ is the total number of available canisters, and $p_i$ is the heat output of the $i$-th fuel element. 
Let us introduce additional variables for indication of fuel element location and canister usage as following:
\begin{align}
	x_{ij} &=
	\begin{cases}
		1, & \text{if $i$-th fuel element is in $j$-th canister}, \\
		0, & \text{otherwise};
	\end{cases} \\
	y_{j} &=
	\begin{cases}
		1, & \text{if $j$-th canister is being used}, \\
		0, & \text{otherwise}.
	\end{cases}
\end{align}

Then, one may formulate optimization problem in the following way: 
\begin{align}
	&M=\sum_{j=1}^{m} y_{j} \rightarrow \min, \label{cnst:0}\\
	& \quad \text{such that} \nonumber \\ 
	&\sum_{i=1}^{n} p_{i} x_{i j} \leq P_{\max } \quad \forall j \in\{1, \ldots, m\}, \label{cnst:1}\\
	&\sum_{j=1}^{m} x_{i j}=1 \quad \forall i \in\{1, \ldots, n\}, \label{cnst:2} \\
	&\sum_{i=1}^{n} x_{i j} \geq N_{\min } y_{j} \quad \forall j \in\{1, \ldots, m\}, \label{cnst:3}\\
	&x_{i j} \leq y_{j} \quad \forall i \in\{1, \ldots, n\},  \forall j \in\{1, \ldots, m\}, \label{cnst:4}
\end{align}
where condition~(\ref{cnst:1}) restricts the maximum heat output per one canister, 
condition~(\ref{cnst:2}) implies that every fuel element placed only in one canister, 
condition~(\ref{cnst:3}) stands for minimal filling of every used canister, 
and condition~(\ref{cnst:4}) binds the variables $x_{i j}$ and $y_j$ so that the placement of the fuel elements matches the vector of the used canisters.
    
The main step in solving an optimization problem using quantum and quantum-inspired annealing is to map the problem of interest to the energy Hamiltonian, 
so the quantum device could find the ground state that corresponds to the optimum value of the objective function. 
The natural way of mathematical description of a quantum annealer is the Ising spin Hamiltonian that can be transformed into QUBO problem in a straightforward way. 
We are to formulate mapping of the CF problem into QUBO form. 
In general, a QUBO problem may be formulated using matrix notation as following:
\begin{equation}
    z^T Q z \to  \min, 
\end{equation}
where $z$ is the vector of binary decision variables and $Q$ is a square symmetric matrix of constants. 

It is necessary to include optimization constraints by adding penalty terms to the objective function. 
Let us represent optimization constraints~(\ref{cnst:1})--(\ref{cnst:4}) in the QUBO form. 
Constraint~\eqref{cnst:1} can be represented as
\begin{equation} \label{qubo:1}
	\mathcal{H}_1 = \sum_{j=1}^{m}\left(\sum_{i=1}^{n} p_{i} x_{ij}+\sum_{l=0}^{s-1} 2^{l} a_{l j}-P_{\max}\right)^{2},
\end{equation}
where $s=\lceil\log _{2} P_{\max }\rceil$ and $a_{lj}$ denotes auxiliary binary variables, 
which are required to represent~\eqref{cnst:1} in the form of equality: $\sum_{l=0}^{s} 2^{l} a_{l j}$ is certain non-negative integer that corresponds to a difference between $P_{\max}$ and $\sum_{i=1}^{n} p_{i} x_{ij}$.
Constraints~\eqref{cnst:2} and \eqref{cnst:3} take the following form:
\begin{equation} \label{qubo:2}
	\mathcal{H}_2 = \sum_{i=1}^{n}\left(\sum_{j=1}^{m} x_{i j}-1\right)^{2},
\end{equation}
and
\begin{equation} \label{qubo:3}
	\mathcal{H}_3 = \sum_{j=1}^{m}\left(\sum_{i=1}^{n} x_{ij}-\sum_{l=0}^{k-1} 2^{l} b_{l j}-N_{\min}y_j\right)^{2},
\end{equation}
correspondingly. 
Here
$k=\lceil\log_{2} N_{\min}\rceil$ and $b_{l j}$ are another auxiliary binary variables used to represent non-negative difference $\sum_{l=0}^{k} 2^{l} b_{l j}$ between $\sum_{i=1}^{n} x_{ij}$ and $N_{\min}y_j$. 
The final constraint~\eqref{cnst:4} takes the form
\begin{equation} \label{qubo:4}
	\mathcal{H}_4 = \sum_{i=1}^n\sum_{j=1}^m \left(x_{i j}-x_{i j} y_{j}\right)
\end{equation}

The problem Hamiltonian then consist of two main components:
\begin{equation}
    \mathcal{H} = A \sum_{j=1}^{m}{y_j} + B \sum_{r=1}^4 \mathcal{H}_r
\end{equation}
where $A$ is a positive constant and $B$ stands for a positive penalty value.
Parameters $A$ and $B$ should be set manually, using the following criteria.
Penalty value should be high enough to keep final solution from violating constraints. 
At the same time, too large penalty value may overwhelm the objective function so it becomes hard to distinguish solutions of different quality.
Therefore, the solution of the optimization problem requires finding optimal values of the variables $x_{ij}, y_{j}, a_{lj}$ and $b_{lj}$. 
More details about the total number of binary variables may be found in Subsec.~\ref{app:dataset}.

One may transform QUBO problem into Ising Hamiltonian using following approach:
\begin{equation}
	z_i = \frac{\sigma^{Z}_i + 1}{2} \in\{0,1\},
\end{equation}
where $\sigma_i^{Z} = \pm 1$ and vector $z$ contains all optimized variables $x_{ij}, y_{j}, a_{lj}$, and $b_{lj}$.

\section{Benchmarking procedure}\label{sec:bench}

In order to evaluate the feasibility of the proposed scheme of solving the SNF problem in the QUBO form, 
we conduct a comparison of existing quantum annealing device and quantum-inspired annealing simulator as instruments to solving CF combinatorial problem.

\subsection{Generating of synthetic dataset}\label{app:dataset}

We first prepare a synthetic dataset of 80 problem instances with number of fuel elements ranging from 3 to 10 and the maximum number of canisters equal to 3. 
For each problem size, 10 different cell configurations with various heat output are prepared (plus single trivial case with 2 elements). 
The optimal allocation for all instances is known in advance and requires at most 2 canisters. 

The general idea of dataset is to create problem with minimal possible QUBO sizes for guaranteed best solution achievement by annealers. 
Using notation from Eqs.(\ref{qubo:1})--(\ref{qubo:4}), the total number of logical transformation variables may be represented as follows:
\begin{equation}
	D = m \left(1 + n + s + k \right),
\end{equation}
where $m$ is the available number of canisters, $n$ is the number of fuel elements, $s=\lceil\log _{2} P_{\max }\rceil$ and $k=\lceil\log_{2} N_{\min}\rceil$ is the number of auxiliary variables used in equalities~(\ref{qubo:1}) and~(\ref{qubo:3}), correspondingly.
As a result, the smallest-size problem with $m=2, n=2, s=2, k=0$ requires 10 logical variables, we mark it as a trivial case (see Fig.~\ref{fig:comparison}). 

We use only fixed configurations, where the optimal number of canisters is 2 and the feasible solution without constraint violation can have 3 canisters, in other words, we restrict problem samples to cases, where $m=3$ and $M=2$. 
The maximum capacities of canisters are equal with limit $2^{s=4}-1$, where $k$ is also fixed and equals zero, since we do not use minimum elements constraint when $N_{\min}=1$. 
This compression allows us to compute model problems on present quantum hardware and evaluate the dependence of problem size and performance. 
Tasks with the same elements quantity are also have identical QUBO size for avoiding additional deviation in data; see Table~\ref{tab:dataset}.
We use public access to D-Wave 5000 Advantage system with Pegasus topology processor~\cite{Boothby2020} to run our quantum annealing algorithm. 

\begin{table}[ht]
\begin{tabular}{|c|c|c|c|c|}
\hline
\multirow{2}{*}{\begin{tabular}[c]{@{}c@{}}\vspace{-0.10cm} Elements\\ quantity \end{tabular}} &
  \multirow{3}{*}{QUBO size} &
  \multicolumn{3}{c|}{Physical qubits} \\ \cline{3-5} 
 &
   &
  \multirow{2}{*}{\begin{tabular}[c]{@{}c@{}}\vspace{-0.10cm} Number \\ of qubits\end{tabular}} &
  \multirow{2}{*}{\begin{tabular}[c]{@{}c@{}}\vspace{-0.10cm} Heuristic, \\ mean\end{tabular}} &
  \multirow{2}{*}{\begin{tabular}[c]{@{}c@{}}\vspace{-0.10cm} Heuristic, \\ std \end{tabular}} \\
   &    &     &       &      \\ \hline
2  & 10 & 20  & 18.4  & 2.3  \\ \hline
3  & 21 & 80  & 64.0  & 3.5  \\ \hline
4  & 24 & 90  & 85.0  & 1.6  \\ \hline
5  & 27 & 102 & 103.8 & 8.6  \\ \hline
6  & 33 & 157 & 153.6 & 5.5  \\ \hline
7  & 36 & 170 & 185.6 & 7.6  \\ \hline
8  & 39 & 185 & 226.6 & 15.6 \\ \hline
9  & 42 & 246 & 246.0 & 10.6 \\ \hline
10 & 45 & 262 & 280.0 & 5.3  \\ \hline
\end{tabular}
\caption{Synthetic dataset scheme. Heuristic embeddings on physical qubits of D-Wave device were found via D-Wave Leap SDK for 5 different random samples.}
\label{tab:dataset}
\end{table}

\subsection{Benchmarking}

\begin{figure*}[htp]
	\includegraphics[width=1\linewidth]{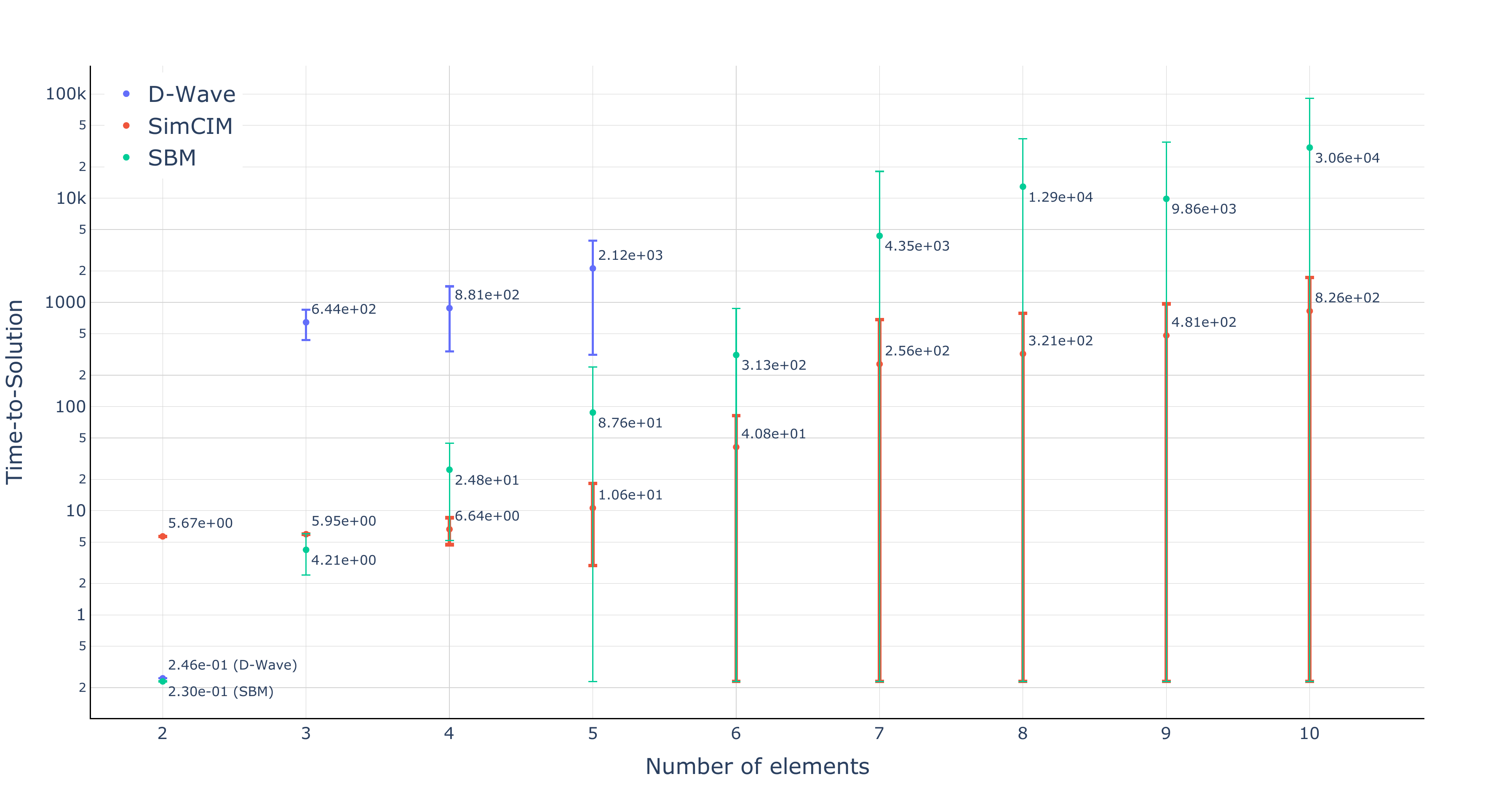}
	\caption{Comparison of the performance of quantum and quantum-inspired methods for bin packing problem based on synthetic data: 
	we compare TTS (mean and standard deviation) for quantum device D-Wave and two quantum-inspired optimization algorithms, SimCIM and SBM.}
	\label{fig:comparison}
\end{figure*}

Each problem instance has been further transformed into the QUBO matrix and run through both quantum and quantum-inspired instruments, specifically, the D-Wave quantum annealer and two quantum-inspired algorithms 
(SimCIM~\cite{Tiunov2019} and SBM~\cite{Goto2019}).
SimCIM algorithm~\cite{Tiunov2019} is based on the method of efficient simulation of Coherent Ising Machine~\cite{Yamamoto2017} using classical computer. 
As it has been shown, SimCIM outperforms Coherent Ising Machine in terms of samples quality and speed of computation, and that is why it has been chosen as a benchmarking tool for comparative analysis.

Simulated Bifurcation algorithm (SBM)~\cite{Goto2019} is a heuristic algorithm for combinatorial optimization. 
Its workflow is inspired by quantum bifurcation machine~\cite{Goto2016} that is based on nonlinear oscillators and implements quantum adiabatic algorithm for solving optimization problems.

We run D-Wave experiments in a pure quantum mode using the Advantage chip featuring 5000 qubits with 15-way connectivity. 
In order to embed QUBO problem into physical qubit layout we utilized clique embedding supported by D-Wave Leap SDK. 
SimCIM was run on Xeon E3-1230v5 4x3,4GHz, 16 GB DDR4, GeForce GTX 1080. 
Comparison results are shown in Fig.~\ref{fig:comparison}.

\subsection{Analysis}

As a figure of merit in our benchmarking procedure, we use time-to-solution (TTS). 
The TTS means a time that is needed to a heuristic algorithm to find the solution (ground state energy) with 99\% success probability.
It is given by
\begin{equation}
	{\rm TTS} = t_a R_{99},
\end{equation}
where $t_a$ is the annealing time (default value of $t_a$ for D-Wave is $20 \mu s$), and $R_{99}$ stands for the number of repetition that is needed for the desired success probability~\cite{Katzgraber2019,Botter2020}.
It can be calculated as follows: 
\begin{equation}\label{r99}
	R_{99} = \frac{\log(1-0.99)}{\log(1-\theta)},
\end{equation}
where $\theta$ is the estimated success probability of each run.

All tasks were grouped by fuel elements quantity for demonstrating results (see Fig.~\ref{fig:comparison}).
We note that the D-Wave annealer shows good result in problem solving in the small-size cases (2 possible canisters and 2 elements), while optimal solution was not found for 6 and more elements problem. 
The standard deviation of TTS is significantly increase with elements quantity for all methods. This is caused by an exponential growth of the space of possible solutions, leading to a decrease in the probability of finding the optimal solution. As a result, the annealing process often terminates at a suboptimal point instead of the ground state. This is especially true for complex problems that require a large number of variables to be taken into consideration. 

While the main obstacle of quantum-inspired optimization methods is complexity and the size of the space of possible solutions, for quantum annealing a very important parameter is the gap between the ground state and the first excited state of the Hamiltonian. The smaller the gap, the slower the adiabatic evolution of the quantum system should proceed in order to stay in the ground state. However, a long evolution time increases the influence of quantum decoherence and can lead to incorrect solutions. 

\section{Conclusion}\label{sec:conclusion}

Quantum computing is a promising technique for solving combinatorial optimization problems.
In our work, we have demonstrated the potential of quantum and quantum-inspired tools to solve computational problems of the minimum bin packing problem, which is formulated as the problem of atomic energy industry, 
As a target problem, we have chosen the optimization of spent nuclear fuel filling in canisters for the deep repository. 
The CF problem has been formulated in a QUBO matrix form (see Eqs.~(\ref{qubo:1})--(\ref{qubo:4})) and was solved using existing quantum annealer and quantum-inspired annealing algorithms. 
We note that the current development level of quantum computing devices does not allow to solve large-scale practical problems, however, it is possible to scale a size of the problem for the next generation of quantum computers.
Moreover, such research helps to identify practical-oriented tasks that may be solved more efficiently by quantum computing. 

\section*{Acknowledgements}

We acknowledge use of the D-Wave quantum annealer and Toshiba Simulated Bifurcation Machine for this work; the views expressed are those of the authors and do not reflect the official policy or position of D-Wave and Toshiba.
The results of the present paper are based on the data that have been collected during the availability of the D-Wave quantum annealer and the Simulated Bifurcation algorithm.
This work was supported by Russian Science Foundation (19-71-10092).

\begin{table}[t]
\begin{tabular}{|c|c|c|}
\hline
Annealing time, $\mu s$ & Success probability & TTS, $\mu s$ \\ \hline
20                  & 0.277               & 284      \\ \hline
40                  & 0.303               & 511      \\ \hline
80                  & 0.343               & 878      \\ \hline
\end{tabular}
\caption{Dependence of success probability and TTS for trivial case (2 fuel elements) for D-Wave device.}
\label{tab:annealing-time}
\end{table}

\section*{Appendix}

In our benchmarking procedure, we use various accessible parameters (particularly, annealing time, embedding type and chain strength) of the quantum hardware on the final solution quality. 
For the simplest task, we have observed that increasing annealing time gives a better success probability (see Table~\ref{tab:annealing-time}), but in the same time TTS is getting worse, so annealing time was set to 20 $\mu$s (the. default value). 
The number of annealing runs is set to $10^4$ (maximum possible value). 
Comparing different embeddings (see Table~\ref{tab:dataset}), we analyze deviation of physical qubits number on five different random samples and decided that for better experiment performance using stable stable clique embedding is more preferable. 
According to Ref.~\cite{Botter2020} we choose custom optimized chain strength instead of other variants such as maximum absolute value in QUBO. 
Thereby we have used mostly the standard configuration of the D-Wave processor during our experiments, so we do not have any specific requirements on the weights/couplers in the model. 
We use only pure quantum regime to obtain solution for each task.

\bibliography{bibliography.bib}

\end{document}